\documentclass[onecolumn,superscriptaddress,aps,prd,nofootinbib]{revtex4-2}
\usepackage{bm,latexsym,amsfonts,mathrsfs}
\usepackage{amsmath,amssymb}
\usepackage[dvipdfmx]{graphicx}
\usepackage[hang,small,bf]{caption}
\usepackage[subrefformat=parens]{subcaption}
\usepackage{ascmac}
\usepackage{physics}
\usepackage{color}
\usepackage{braket}
\usepackage[dvipsnames]{xcolor}
\definecolor{red}{rgb}{0.9, 0,0}
\definecolor{cerulean}{rgb}{0., 0.42,0.9}
\definecolor{navy}{rgb}{0.05, 0.05,0.8}

\usepackage{setspace}
\usepackage[normalem]{ulem}

\newcommand{\simgt}{\lower.5ex\hbox{$\; \buildrel > \over \sim \;$}}
\newcommand{\simlt}{\lower.5ex\hbox{$\; \buildrel < \over \sim \;$}}
\usepackage{comment}

\usepackage[colorlinks]{hyperref}
\makeatletter
\providecommand{\href}[2]{#2}        
\providecommand{\href@noop}[2]{#2}   
\providecommand{\Eprint}[2][]{#2}    
\makeatother
\begin{document}

\title{Momentum Squeezed State Realized via Optimal Filtering in Optomechanics: Implications for Gravity-Induced Entanglement}
\author{Ryotaro Fukuzumi}
\email{fukuzumi.ryotaro.709@s.kyushu-u.ac.jp}
\affiliation{\small\it Department of Physics, Kyushu University, 744 Motooka, Nishi-Ku, Fukuoka 819-0395, Japan}
\author{Kosei Hatakeyama}
\email{hatakeyama.kosei.103@s.kyushu-u.ac.jp}
\affiliation{\small\it Department of Physics, Kyushu University, 744 Motooka, Nishi-Ku, Fukuoka 819-0395, Japan}
\author{Daisuke Miki}
\email{dmiki@caltech.edu}
\affiliation{\small\it The Division of Physics, Mathematics and Astronomy, California Institute of Technology, Pasadena, CA 91125, USA}
\author{Kazuhiro Yamamoto}
\email{yamamoto@phys.kyushu-u.ac.jp}
\affiliation{\small\it Department of Physics, Kyushu University, 744 Motooka, Nishi-Ku, Fukuoka 819-0395, Japan}
\affiliation{\small\it Quantum and Spacetime Research Institute, Kyushu University, 744 Motooka, Nishi-Ku, Fukuoka 819-0395 Japan}
\date{\today}

\begin{abstract}
We analyze the conditional quantum state of a mechanical mirror in an optomechanical system subject to continuous measurement, feedback control, and quantum filtering. 
We identify a parameter regime in which the mirror exhibits momentum squeezing beyond the standard quantum limit, achieved through an appropriate choice of the homodyne detection angle. 
In this regime, we show that optimal filtering effectively realizes a free-particle-like conditional state. 
When this mechanism is applied to a configuration consisting of two optomechanical systems, the resulting momentum squeezing significantly enhances the signal of gravity-induced entanglement (GIE). 
This enhancement arises because the momentum squeezing not only amplifies the distinction between the common and differential modes, but also, in the high-purity regime, increases the position uncertainty in accordance with the uncertainty principle, thereby enlarging the spatial extent of the quantum superposition. 
Our results provide new insights into experimental strategies for probing the quantum nature of gravity using optomechanical platforms. 

\end{abstract}

\maketitle
\section{Introduction}
Cavity optomechanical systems, consisting of an optical cavity mode coupled to a mechanical oscillator, provide a promising and highly controllable platform for realizing and probing macroscopic quantum states~\cite{Chen2013,Aspelmeyer2014,Bowen}. 
In particular, it has been demonstrated both theoretically and experimentally that continuous quantum measurement, when combined with appropriate feedback and quantum control, enables the cooling of a mechanical oscillator close to or even into its quantum ground state~\cite{Whittle2021,Matsumoto}. 
Beyond ground-state preparation, this measurement-based approach allows one to engineer nonclassical conditional states of motion, such as squeezed and entangled states, in regimes that would otherwise be inaccessible due to thermal noise and decoherence. 
Utilizing these techniques, the authors of Refs.~\cite{Helge2008,Miki2023} demonstrated the potential for generating and stabilizing entanglement between spatially separated or distinct macroscopic mechanical objects under continuous measurement, where the quantum optimal filter plays a crucial role in extracting the maximum amount of information from the measurement record and in suppressing classical noise. 
Such conditional entanglement schemes provide a powerful framework for exploring quantum correlations at increasingly larger mass and length scales, pushing the boundary between the quantum and classical worlds. 

More recently, significant attention has been directed toward the possibility of observing gravity-induced entanglement (GIE) \cite{Bose2017,Marletto2017,Krisnanda2019}. 
The central idea is that if gravity itself acts as a quantum mediator, then two massive objects interacting solely through their mutual gravitational attraction can become entangled \cite{Feynman1995}. 
Observation of such entanglement would provide strong evidence that gravity possesses the intrinsically quantum nature, offering a novel and experimentally accessible test of quantum gravity in the low-energy, non-relativistic regime \cite{Krisnanda2019}. 
Recent theoretical studies suggest that gravity-induced entanglement may be experimentally accessible in optomechanical systems, highlighting their potential as a platform for testing the quantum nature of gravity \cite{Miao2020,Datta2021,Miki2024a,Miki2024b,Matsumoto2025,Hatakeyama2025}. 

In the present paper, we investigate the quantum state of an optomechanical system subject to continuous measurement, feedback control, and optimal filtering. 
We identify a parameter regime in which the conditional state of the mechanical mirror exhibits strong momentum squeezing, even though the measurement readout corresponds to the oscillator position. 
In this regime, optimal filtering yields a free-particle-like conditional state, which can be interpreted as enabling backaction-evading momentum detection. 
We analytically investigate the conditions under which such a momentum-squeezed state emerges and clarify the underlying physical mechanism, highlighting the essential role played by quantum optimal filtering. 
Furthermore, we demonstrate that the momentum-squeezed state significantly enhances GIE in an optomechanical setup with two mirrors, as proposed in Ref.~\cite{Miki2024b}. 
This enhancement may be crucial for experimental tests aimed at verifying the quantum nature of gravity. 

The present paper is organized as follows. 
In Sec.~\ref{Formulas}, we introduce the fundamental formalism of an optomechanical system and describe how the conditional state of the mechanical mirror is inferred from continuous measurements of the mirror position coupled to the optical field. 
We also discuss how the application of an optimal filter, designed to minimize the conditional uncertainty, leads to an effective shift in the resonance frequency and damping rate. 
In Sec.~\ref{Momentum squeezing}, we analyze the conditional variances and identify the parameter region in which a momentum-squeezed state emerges, characterized by a strong reduction of the momentum uncertainty. 
We further clarify the conditions under which this regime is realized. 
In Sec.~\ref{Gravity-induced entanglement}, we demonstrate that gravity-induced entanglement (GIE) is significantly enhanced in the parameter region identified in the preceding sections. 
Sec.~\ref{Conclusions} is devoted to a summary and concluding remarks. 
In Appendix~\ref{Appendix:feedback}, we examine the effect of feedback control. 
We show that the conditional variances contain no contribution from the feedback control and that measurement-based feedback implemented with a high-pass filter does not affect the conditional variances. 

\section{Formulas}
\label{Formulas}
We consider a cavity optomechanical system consisting of a mechanical mirror described by the canonical operators $(\hat{Q},\hat{P})$, with mass $m$ and bare mechanical frequency $\Omega$, coupled to an optical cavity mode $(\hat{a},\hat{a}^\dagger)$ with resonance frequency $\omega_c$ and cavity length $\ell$. 
The system Hamiltonian is given by 
\begin{align}
    \hat{H}
    = \frac{\hat{P}^2}{2m}
    + \frac{1}{2} m \Omega^2 \hat{Q}^2
    + \frac{\hbar \omega_c}{1+\hat Q/\ell}\hat{a}^\dagger \hat{a}
    + i \hbar E \left( e^{i\omega_L t} \hat{a}^\dagger - e^{-i\omega_L t} \hat{a} \right), 
\end{align}
where the canonical commutation relation $[\hat{Q},\hat{P}] = i\hbar$ is assumed. 
Here, $\omega_L$ is the laser frequency, and the input laser amplitude is $E = \sqrt{P_\mathrm{in}\kappa / (\hbar\omega_L)}$, with $P_\mathrm{in}$ denoting the input laser power and $\kappa$ the cavity decay rate. 
We here expand the mechanical position operator into a classical mean value and a quantum fluctuation as $\hat{Q} \rightarrow \bar{Q} + \hat{Q}$. 
Moving to the rotating frame at the laser frequency, the Hamiltonian can be written as 
\begin{align}
    \hat{H}
    = \frac{\hbar\Omega}{4}(\hat{q}^2 + \hat{p}^2)
    - \hbar \Delta \hat{a}^\dagger \hat{a}
    + \hbar g_0 \hat{q} \hat{a}^\dagger \hat{a}, 
\end{align}
where the dimensionless mechanical quadratures $\hat{q}=\sqrt{2m\Omega/\hbar}\hat{Q}$ and $\hat{p}=\sqrt{2/m\hbar\Omega}\hat{P}$ satisfy $[\hat{q},\hat{p}] = 2i$. 
The
optomechanical coupling strength is given by $g_0 = (\omega_c/\ell)\sqrt{\hbar/(2m\Omega)}$,
and $\Delta=\omega_L-\omega_c(1-\bar{Q}/\ell)$ is the detuning parameter. 

Assuming a large intracavity photon number, we linearize the system by expanding the cavity field operator as $\hat{a} \rightarrow \bar{a} + \hat{a}$, where $\bar{a}$ denotes the classical steady-state amplitude. 
Introducing the amplitude and phase quadratures of the optical field, $\hat{x} = \hat{a} + \hat{a}^\dagger$ and $\hat{y} = -i(\hat{a} - \hat{a}^\dagger)$, which satisfy $[\hat{x},\hat{y}] = 2i$, the linearized Langevin equations read 
\begin{align}
    \dot{\hat{q}} &= \Omega \hat{p}, \qquad
    \dot{\hat{p}} = -\Omega \hat{q} - 2 g \hat{x} - \Gamma \hat{p}
    + \sqrt{2\Gamma}\hat{p}_\mathrm{in}, \\
    \dot{\hat{x}} &= -\Delta \hat{y} - \frac{\kappa}{2}\hat{x}
    + \sqrt{\kappa}\hat{x}_\mathrm{in}, \\
    \dot{\hat{y}} &= \Delta \hat{x} - 2 g \hat{q}
    - \frac{\kappa}{2}\hat{y}
    + \sqrt{\kappa}\hat{y}_\mathrm{in}, 
\end{align}
where the linearized optomechanical coupling is 
\begin{align}
    g = g_0 \abs{\bar{a}}
    = \sqrt{\frac{\omega_c P_\mathrm{in}\kappa}
    {m\Omega\ell^2(\kappa^2 + 4\Delta^2)}}. 
\end{align}
$\Gamma$ denotes the mechanical dissipation rate, and the mechanical thermal noise operator $\hat{p}_\mathrm{in}$ has zero mean and satisfies 
\begin{align}
    \frac{1}{2}\langle\{\hat{p}_\mathrm{in}(t), \hat{p}_\mathrm{in}^\dagger(t')\}\rangle
    = \left(\frac{2k_B T}{\hbar\Omega} + 1\right)\delta(t-t'),
\end{align}
where $k_B$ is the Boltzmann constant and $T$ the temperature. 
We assume vacuum input noise for the optical fields, with correlations 
\begin{align}
    \frac{1}{2}\langle\{\hat{x}_\mathrm{in}(t), \hat{x}_\mathrm{in}^\dagger(t')\}\rangle
    = \frac{1}{2}\langle\{\hat{y}_\mathrm{in}(t), \hat{y}_\mathrm{in}^\dagger(t')\}\rangle
    \simeq \delta(t-t'), \qquad
    \langle\{\hat{x}_\mathrm{in}(t), \hat{y}_\mathrm{in}^\dagger(t')\}\rangle = 0. 
\end{align}

Using the input–output relations $\hat{X} = \hat{x}_\mathrm{in} - \sqrt{\kappa}\hat{x}$ and $\hat{Y} = \hat{y}_\mathrm{in} - \sqrt{\kappa}\hat{y}$ \cite{Gardiner,Aspelmeyer2014}, and assuming the adiabatic limit $\kappa \gg \omega_m$, the equations of motion for the mechanical system and the output quadratures reduce to 
\begin{align}
    \label{Ffbt}
    &\dot{\hat{q}}' = \omega_m \hat{p}', ~~~
    \dot{\hat{p}}' = -\omega_m \hat{q}' - \Gamma \hat{p}'
    + \hat{w}(t) + F_\mathrm{fb}(t), \\
    \label{defw}
    &\hat{w} (t)=
    \sqrt{2\Gamma}\hat{p}_\mathrm{in}'
    - \frac{4 g_m \kappa \sqrt{\kappa}}{\kappa^2 + 4\Delta^2}\hat{x}_\mathrm{in}
    + \frac{8 g_m \Delta \sqrt{\kappa}}{\kappa^2 + 4\Delta^2}\hat{y}_\mathrm{in}, 
\end{align}
together with the output fields 
\begin{align}
    \hat{X} &=
    - \frac{8 g_m \Delta \sqrt{\kappa}}{\kappa^2 + 4\Delta^2}\hat{q}'
    - \frac{\kappa^2 - 4\Delta^2}{\kappa^2 + 4\Delta^2}\hat{x}_\mathrm{in}
    + \frac{4\Delta\kappa}{\kappa^2 + 4\Delta^2}\hat{y}_\mathrm{in}, \\
    \hat{Y} &=
    \frac{4 g_m \kappa \sqrt{\kappa}}{\kappa^2 + 4\Delta^2}\hat{q}'
    - \frac{4\Delta\kappa}{\kappa^2 + 4\Delta^2}\hat{x}_\mathrm{in}
    - \frac{\kappa^2 - 4\Delta^2}{\kappa^2 + 4\Delta^2}\hat{y}_\mathrm{in}.
\end{align}
Here we have introduced the rescaled quadratures and effective parameters 
\begin{align}
    \label{rescaled quadratures}
    \hat{q}' &= \hat{q}\sqrt{\frac{\omega_m}{\Omega}}, \qquad
    \hat{p}' = \hat{p}\sqrt{\frac{\Omega}{\omega_m}}, \\
    g_m & = g\sqrt{\frac{\Omega}{\omega_m}}, \qquad
    \omega_m = \sqrt{\Omega^2
    + \Omega\frac{16 g^2 \Delta}{\kappa^2 + 4\Delta^2}},
\end{align}
where $\omega_m$ is the optically induced mechanical frequency due to the optical spring effect \cite{Buonanno}. 
In the following, we simply denote $(\hat{q}',\hat{p}')$ as $(\hat{q},\hat{p})$. 

In Eq.~\eqref{Ffbt}, we have added $F_\mathrm{fb}(t)$, which represents a feedback force applied to the mechanical oscillator based on the measurement outcome \cite{Genes2008}. 
Although this term modifies the mechanical damping and contributes additional fluctuations in the unconditional description,
the results presented below do not depend on the feedback parameters. 
This is because $\hat F_\mathrm{fb}$ is a known control input constructed from the measurement outcome and does not affect the conditional covariance matrix.
As shown
explicitly in Appendix~\ref{Appendix:feedback}, the feedback-induced terms can be eliminated by applying an appropriate optimal filter.

Hence, the subsequent analysis remains valid even in the presence of feedback. 

\begin{figure}[t]
    \centering
    \includegraphics[width=100mm]{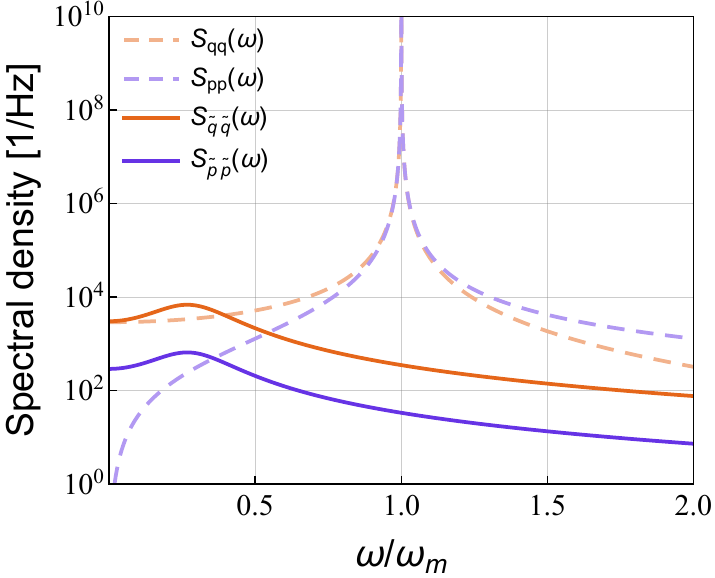}
    \caption{
            Spectral densities of the mechanical position $q$ and momentum $p$ without and with causal Wiener filtering. 
            The parameters are chosen as
            $m = 100~\mathrm{g}$,
            $\ell = 10~\mathrm{cm}$,
            $\Omega/2\pi = 10^{-3}~\mathrm{Hz}$,
            $\kappa/2\pi = 10^{8}~\mathrm{Hz}$,
            $\omega_c/2\pi = 2.818\times10^{14}~\mathrm{Hz}$,
            $\Gamma/2\pi = 10^{-18}~\mathrm{Hz}$,
            $P_\mathrm{in} = 10^{-5}~\mathrm{W}$,
            $T = 1~\mathrm{K}$,
            $\Delta = 0$, and
            $\theta = -\pi/60$,
            which yield $\omega_\theta/\omega_m \simeq 0.31$ and
            $\gamma_\theta/2\pi \simeq 3.0\times10^{-4}~\mathrm{Hz}$. 
            Without filtering, the spectral densities of $(q,p)$ exhibit a peak at the mechanical resonance frequency $\omega_m$ (dashed curves). 
            In contrast, when causal Wiener filtering is applied, the spectral densities of the filtered variables $(\tilde{q},\tilde{p})$ show a shifted peak at $\omega_\theta$ (solid curves). 
            }
    \label{SqqSpp}
\end{figure}

Defining the Fourier transform as $f(\omega) = \int_{-\infty}^{\infty} dt\, f(t)e^{i\omega t}$, the steady-state solutions of the Langevin equations are given by 
\begin{align}
    \hat{q}(\omega) = \frac{\omega_m}{F(\omega)} \hat{w}(\omega), \qquad
    \hat{p}(\omega) = -\frac{i\omega}{F(\omega)} \hat{w}(\omega), 
\end{align}
where $F(\omega) = \omega_m^2 - i\Gamma\omega - \omega^2$ denotes the complex mechanical susceptibility characterized by $\omega_m$ and $\Gamma$, and $\hat{w}(\omega)$ is the Fourier transform of $\hat{w}(t)$. 
The unconditional spectral densities $S_{qq}(\omega)$ and $S_{pp}(\omega)$, which are defined by the variance of $\hat{q}(\omega)$ and $\hat{p}(\omega)$, respectively, exhibit resonance peaks at $\omega = \pm \omega_m$. 
The dashed curves in Fig.~\ref{SqqSpp} show the unconditional spectra for $\omega>0$, clearly displaying the resonance at $\omega_m$. 

Here, we consider a linear measurement of a general output quadrature, 
\begin{align}
    \hat{I}_\theta = \hat{X}\cos\theta + \hat{Y}\sin\theta, 
\end{align}
which corresponds to homodyne detection with a local-oscillator phase $\theta$. 
Throughout this paper, we assume an ideal (perfect-efficiency) homodyne measurement at an arbitrary angle $\theta$. 
Using the input-output relations derived above, the general output quadrature $\hat{I}_\theta$ can be rewritten as 
\begin{align}
    \hat{I}_\theta = c_\theta\hat{q} + \hat{v}_\theta, 
\end{align}
where we define 
\begin{align}
    \label{defc_theta}
    &c_\theta = -\frac{8g_m\Delta\sqrt{\kappa}}{\kappa^2 + 4\Delta^2} \cos\theta + \frac{4g_m\kappa\sqrt{\kappa}}{\kappa^2 + 4\Delta^2}\sin\theta, \\
    \label{defv}
    &\hat{v}_{\theta} = - \frac{(\kappa^2 - 4\Delta^2) \cos\theta + 4\Delta\kappa \sin\theta}{\kappa^2 + 4\Delta^2}\hat{x}_\mathrm{in} + \frac{4\Delta\kappa\cos\theta -(\kappa^2 - 4\Delta^2)\sin\theta}{\kappa^2 + 4\Delta^2}\hat{y}_\mathrm{in}. 
\end{align}

\begin{figure}[t]
    \centering
    \includegraphics[width=100mm]{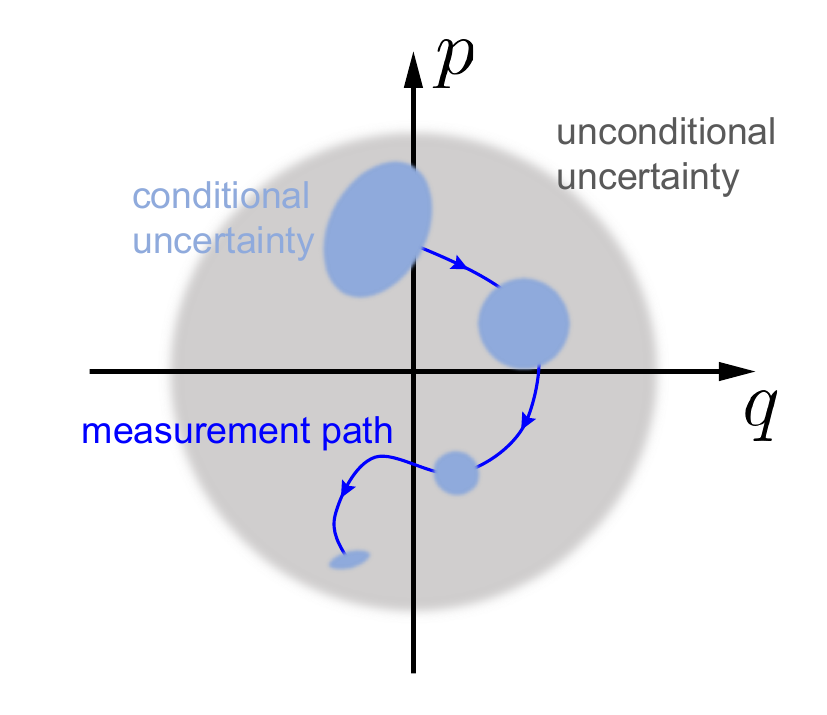}
    \caption{
            Schematic illustration of conditional states in phase space. 
            The blue curve represents the trajectory of the conditional first moments revealed by a continuous measurement process. 
            Information acquired through the measurement reduces the uncertainty, resulting in a blue ellipse that characterizes the conditional uncertainty. 
            The gray circle denotes the unconditional uncertainty obtained by averaging over all possible measurement trajectories. 
            }
    \label{Fig:ConditionalState}
\end{figure}

In the theory of continuous quantum measurement, the conditional state represents the best estimate of the physical state of the system conditioned on the measurement record. 
It evolves according to a stochastic master equation, and its properties—such as the conditional variances—are physically meaningful and experimentally verifiable through ensemble averages. 
The Kalman filter provides such an optimally estimated physical state, as illustrated in Fig.~\ref{Fig:ConditionalState}. 
Owing to the equivalence between Kalman filtering and Wiener filtering in equilibrium, the conditional state obtained with the corresponding conditional variances and correlations is a genuine physical state. 
As demonstrated in quantum feedback and quantum trajectory experiments, this state can be experimentally inferred from the statistics of the measured homodyne photocurrent. 

In the present paper, we adopt a quantum causal Wiener filtering approach under continuous measurement of the output quadrature $\hat{I}_\theta$. 
Wiener filtering provides a systematic method for constructing a conditional quantum state with minimal uncertainty by optimally processing the information acquired through a continuous measurement record \cite{Meng2020}. 
In practice, the Wiener filter is determined by minimizing the variances of the filtered variables, 
\begin{align}
    \tilde{q}(\omega) = \hat{q}(\omega) - H_q(\omega)\hat{I}_\theta(\omega), \qquad 
    \tilde{p}(\omega) = \hat{p}(\omega) - H_p(\omega)\hat{I}_\theta(\omega), 
\end{align}
where $H_q(\omega)$ and $H_p(\omega)$ are causal filter functions. 
It is well known that the optimal Wiener filters are given by Refs.~\cite{Meng2020,Matsumoto,Hatakeyama2025} as 
\begin{align}
    H_q(\omega) &= \frac{1}{S_{II}^\mathrm{C}(\omega)}\Bigg[\frac{S_{qI}(\omega)}{S_{II}^\mathrm{NC}(\omega)}\Bigg]_+, \\
    H_p(\omega) &= \frac{1}{S_{II}^\mathrm{C}(\omega)}\Bigg[\frac{S_{pI}(\omega)}{S_{II}^\mathrm{NC}(\omega)}\Bigg]_+, 
\end{align}
where $[Z(\omega)]_+$ denotes the causal component of $Z(\omega)$, and $S_{II}^\mathrm{C}(\omega)$ and $S_{II}^\mathrm{NC}(\omega)$ represent the causal and noncausal parts of the spectral density $S_{II}(\omega)=S_{II}^\mathrm{C}(\omega)S_{II}^\mathrm{NC}(\omega)$, respectively.
The explicit expressions of the Wiener filters are given by 
\begin{align}
    \label{Hqomega}
    H_q(\omega) &= \frac{1}{c_\theta}\left(1 - \frac{F(\omega)}{F^\prime(\omega)}\right), \\
    \label{Hpomega}
    H_p(\omega) &= -\frac{1}{c_\theta\omega_m}\left[i\omega + (\gamma_\theta - \Gamma - i\omega)\frac{F(\omega)}{F^\prime(\omega)}\right], 
\end{align}
where we have defined $F^\prime(\omega) = \omega_\theta^2 - i\gamma_\theta\omega - \omega^2$ with 
\begin{align}
    \omega_\theta = \omega_m \sqrt[4]{1 + \frac{2\Lambda_\theta}{\omega_m} + \frac{\bar{n}\lambda_\theta}{\omega_m^2}}, \qquad 
    \gamma_\theta = \sqrt{\Gamma^2 - 2\omega_m(\omega_m + \Lambda_\theta) + 2\omega_\theta^2}. 
\end{align}
Here, $\lambda_\theta = c_\theta^2$ and $\Lambda_\theta = c_\theta L_\theta$. 
We note that the parameter $\lambda_\theta$ represents the measurement rate, namely, the inverse timescale required to resolve the zero-point motion.
The quantities $\bar{n}$ and $L_\theta$ are defined through the relations $\braket{\{\hat{w}(t),\hat{w}(t^\prime)\}}/2 = \bar{n}\delta(t-t^\prime)$ and $\braket{\{\hat{w}(t),\hat{v}_\theta(t^\prime)\}}/2 = L_\theta\delta(t-t^\prime)$. 
We also note that Eqs.~\eqref{Hqomega} and \eqref{Hpomega} are the extension of 
Refs.~\cite{Meng2020,Matsumoto,Hatakeyama2025} to a general homodyne angle, and consistent with their results for an appropriate choice of the homodyne angle.

The conditional spectral densities of the mechanical mirror are defined by the variances of the mechanical quadratures conditioned on the measurement outcomes, $\tilde{q}(\omega)$ and $\tilde{p}(\omega)$. 
The conditional spectral densities of $\tilde{q}$ and $\tilde{p}$ are explicitly obtained by 
\begin{align}
    \label{spectrumqq}
    & S_{\tilde{q}\tilde{q}}(\omega) = \frac{1}{\abs{F^\prime(\omega)}^2}\biggl\{\omega_m^2\bar{n} - \frac{2\omega_m(\omega^{2}_\theta - \omega^{2}_m)}{c_\theta}L_\theta + \frac{\omega^2(\gamma_\theta - \Gamma)^2 + (\omega_\theta^2 - \omega_m^2)^2}{c^{2}_\theta}\biggr\}, \\
    \label{spectrumpp}
    &S_{\tilde{p}\tilde{p}}(\omega) = \frac{1}{\abs{F^\prime(\omega)}^2}\biggl\{\qty[(\gamma_\theta - \Gamma)^2 + \omega^2]\bar{n} - \frac{2L_\theta \qty[\omega^2(\omega_\theta^2 - \omega_m^2 - (\gamma_\theta - \Gamma)\Gamma) - \omega_m^2(\gamma_\theta - \Gamma)^2]}{c_\theta\omega_m} \nonumber\\
    &\hspace{15mm} +\frac{\omega^2\qty[(\gamma_\theta - \Gamma)\Gamma - (\omega_\theta^2 - \omega_m^2)]^2 + (\gamma_\theta - \Gamma)^2\omega_m^4}{c_\theta^2\omega_m^2}\biggr\},\\
    \label{spectrumqp}
    &\mathrm{Re}[S_{\tilde q\tilde p}(\omega)] = \frac{\gamma_\theta - \Gamma}{\abs{F^\prime(\omega)}^2} \biggl\{\omega_m\bar{n} - \frac{\omega^2 + (\omega_\theta^2 - 2\omega_m^2)}{c_\theta}L_\theta  - \frac{\qty[\Gamma(\gamma_\theta - \Gamma) - (\omega_\theta^2 - \omega_m^2)]\omega^2 + (\omega_\theta^2 - \omega_m^2)\omega_m^2}{c_\theta^2\omega_m}\biggr\}. 
\end{align}
These conditional spectra exhibit peaks at frequencies $\pm\omega_\theta$. 
The solid curves in Fig.~\ref{SqqSpp} demonstrate $S_{\tilde{q}\tilde{q}}$ and $S_{\tilde{p}\tilde{p}}$. 
This behavior can be understood from the explicit forms of $H_q(\omega)$ and $H_p(\omega)$, together with the fact that $\abs{F^\prime(\omega)}^2$ is minimized at $\omega = \pm\omega_\theta$. 
As a consequence, the effective resonance frequency and dissipation rate of the conditional state can be directly identified from the effective susceptibility $F^\prime(\omega)$. 
Specifically, $\omega_\theta$ and $\gamma_\theta$ correspond to the effective resonance frequency and damping rate induced by the Wiener filtering, respectively. 

The solid and dashed curves in Fig.~\ref{SqqSpp} compare the conditional and unconditional spectral densities. 
While the unconditional spectra exhibit sharp peaks at the mechanical resonance frequency $\omega_m$, the conditional spectra exhibit a moderate enhancement near $\omega_\theta \simeq 0$. 
The parameters used in the calculation, listed in the caption of Fig.~\ref{SqqSpp}, are chosen to realize the momentum-squeezed regime discussed below. 
This qualitative difference in the spectral features originates from the fact that the effective susceptibility of the conditional mechanical state is given by $F^\prime(\omega)$, whereas that of the unconditional state is described by $F(\omega)$. 

The conditional variances and covariance are obtained by integrating the corresponding spectral densities as 
\begin{align}
    V_{qq} &= \frac{1}{2\pi}\int^\infty_{-\infty}d\omega S_{\tilde{q}\tilde{q}}(\omega) = \frac{\gamma_\theta-\Gamma}{\lambda_\theta}, \\
    V_{qp} &= \frac{1}{2\pi}\int^\infty_{-\infty}d\omega \mathrm{Re}\left[S_{\tilde{q}\tilde{p}}(\omega)\right] = V_{qq}\frac{\gamma_\theta - \Gamma}{2\omega_m}, \\
    \label{vpp}
    V_{pp} &= \frac{1}{2\pi}\int^\infty_{-\infty}d\omega S_{\tilde{p}\tilde{p}}(\omega) = V_{qq}\left[\left(\frac{\omega_\theta}{\omega_m}\right)^2 + \frac{\Gamma(\Gamma-\gamma_\theta)}{2\omega_m^2}\right], 
\end{align}
where $V_{qq}$ and $V_{pp}$ denote the conditional position and momentum variances, respectively, and $V_{qp}$ is the conditional position--momentum covariance. 
As shown in Appendix~\ref{Appendix:feedback}, all these components remain unchanged even when considering feedback control.

\section{Momentum squeezing}
\label{Momentum squeezing}
Fig.~\ref{SqqSpp} shows that the maximum value of $S_{\tilde{p}\tilde{p}}$ is significantly smaller than that of $S_{\tilde{q}\tilde{q}}$. 
This implies that the momentum variance $V_{pp}$ is much smaller than the position variance $V_{qq}$, a feature that can be explicitly confirmed by evaluating the conditional variances and covariance. 

We find that the regime $V_{pp} \ll V_{qq}$ emerges in the limit of $\omega_\theta \ll \omega_m$. 
This can be understood as follows. 
The effective dissipation rate is always at least as large as the bare dissipation rate, $\gamma_\theta\ge\Gamma$, and the second term in Eq.~\eqref{vpp} is non-positive. 
Together with the positivity of the variance $V_{pp}>0$, this implies that the momentum variance is bounded from above by the first term in Eq.~\eqref{vpp}. 
Hence, the ratio $\omega_\theta/\omega_m$ determines the ratio $V_{pp}/V_{qq}$. 
In the following, we examine the specific conditions under which $\omega_\theta \ll \omega_m$ is satisfied, leading to the realization of the regime $V_{pp} \ll V_{qq}$. 

To this end, we introduce 
\begin{align}
    \label{defalpha}
    \alpha &= \arctan\frac{2\Delta}{\kappa}, \\
    \xi &= \frac{16 g_m^2\kappa}{\omega_m (\kappa^2 + 4 \Delta^2)} 
    = \frac{4C_m}{Q_m}\cos^2\alpha, 
\end{align}
where $Q_m = \omega_m/\Gamma$ is the mechanical quality factor and $C_m = 4g_m^2/(\Gamma\kappa)$ is the optomechanical cooperativity. 
With these definitions, the quantities $L_\theta$, $\lambda_\theta$, and $\Lambda_\theta$ can be rewritten as $L_\theta = \sqrt{\omega_m\xi}\cos(\theta - \alpha)$, $\lambda_\theta = \omega_m\xi\sin^2(\theta - \alpha)$, and $\Lambda_\theta = \omega_m\xi\sin(\theta - \alpha)\cos(\theta - \alpha)$. 

Using the relation 
\begin{align}
    \left(\frac{\omega_\theta}{\omega_m}\right)^4
    = 1 + \frac{\zeta\xi}{2}
    + \frac{\xi}{2}\sqrt{\zeta^2 + 4}
    \sin\!\left(2\theta - 2\alpha - \arctan\frac{\zeta}{2}\right), 
\end{align}
which can be straightforwardly derived, we find that the ratio $\omega_\theta/\omega_m$ is minimized for the optimized homodyne angle 
\begin{align}
    \label{deftheta}
    \theta = \alpha - \frac{1}{2}\arctan\qty[\frac{2}{\zeta}], 
\end{align}
where $\zeta = \eta + \xi$ denotes the total mirror noise normalized by $\omega_m$. 
Here, $\eta = 2\Gamma(2n_\mathrm{th} + 1)/\omega_m$ represents the normalized thermal noise, while $\xi$ corresponds to the radiation-pressure noise. 

\begin{figure}[t]
    \centering
    \includegraphics[width=90mm]{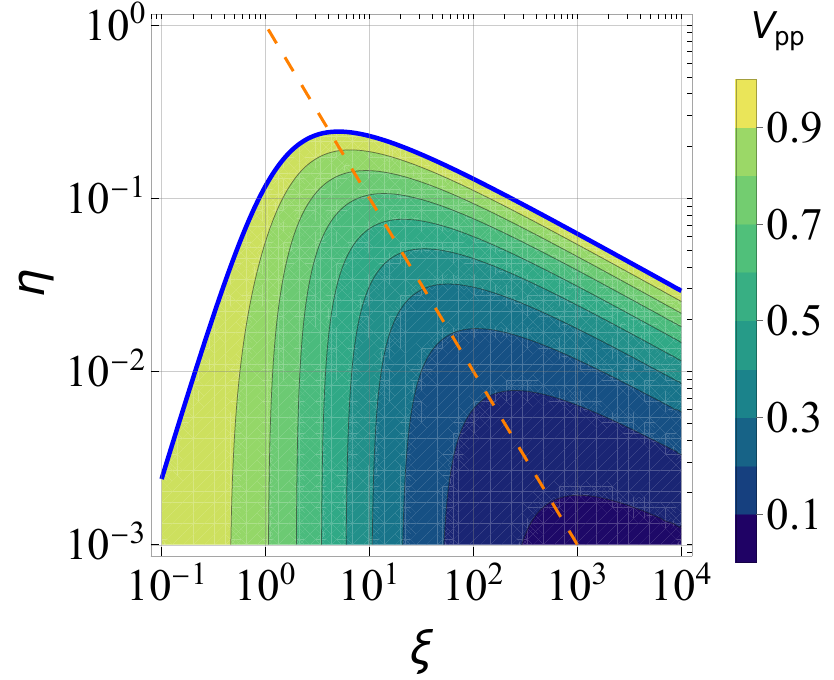}
    \caption{
            Contour plot of $V_{pp}$ obtained under measurement at the optimized homodyne angle given in Eq.~\eqref{deftheta}. 
            The colored region indicates the parameter space in which momentum squeezing occurs, while the blue curve denotes the boundary defined by $V_{pp}=1$. 
            The orange dashed line corresponds to $\eta \xi = 1$, which defines the boundary of the approximation adopted in Eqs.~\eqref{approx:etaxi<1} and \eqref{approx:etaxi>1}. 
            We also note that the approximate formula Eq.~\eqref{etaxiap} is valid for $\eta\ll\xi$ and $2\ll\xi$. 
            }
    \label{thermal-noise}
\end{figure}

For this choice of $\theta$, one finds not only that the momentum variance is always smaller than the position variance, $V_{pp} < V_{qq}$, but also that genuine momentum squeezing, $V_{pp} < 1$, occurs when the thermal noise contribution is sufficiently small, as illustrated in Fig.~\ref{thermal-noise}.
In Fig.~\ref{thermal-noise}, the colored region corresponds to $V_{pp}<1$, while the blue curve marks the boundary $V_{pp}=1$.
In contrast, when thermal noise becomes non-negligible, momentum squeezing is no longer observed.

\begin{figure}[t]
    \centering
    \includegraphics[width=90mm]{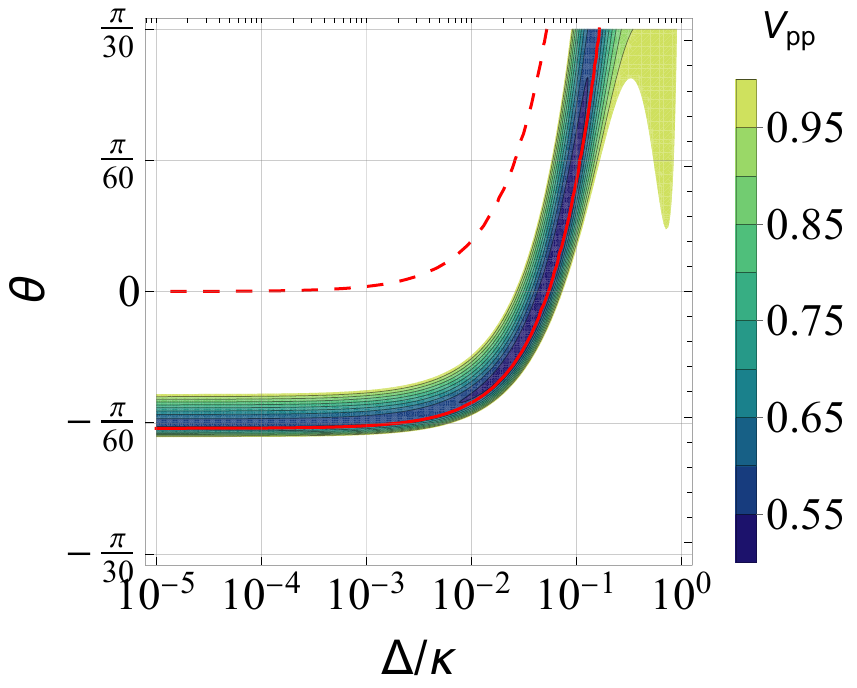}
    \caption{
            Contour plot of $V_{pp}$ in the $(\Delta/\kappa,\,\theta)$ plane. 
            The red solid curve indicates the condition given by Eq.~\eqref{deftheta}, which coincides with the parameter region where momentum squeezing occurs. 
            All parameters are the same as those used in Fig.~\ref{SqqSpp}, except for $\theta$ and $\Delta$. 
            The red dashed curve denotes the zero-measurement-rate condition, $\lambda_\theta = 0$. 
            }
    \label{VppContour}
\end{figure}

Fig.~\ref{VppContour} presents a contour plot of $V_{pp}$ in the $(\Delta/\kappa,\,\theta)$ plane, with all remaining parameters chosen to be identical to those listed in the caption of Fig.~\ref{SqqSpp}. 
The colored region indicates the parameter regime in which $V_{pp}<1$, namely, where momentum squeezing occurs. 
The red solid curve corresponds to the relation given in Eq.~\eqref{deftheta}, while the red dashed curve represents the zero-measurement-rate condition, $\lambda_\theta=0$, given by $\theta=\alpha$. 
It is evident that the red solid curve precisely overlaps with the region where momentum squeezing is observed. 
This clearly indicates that the condition for achieving momentum squeezing is equivalent to minimizing the ratio $\omega_\theta/\omega_m$.
The zero-measurement-rate condition corresponds to a regime in which the measurement outcomes contain no information about the mirror position. 
In this case, the mechanical state cannot be conditioned on the measurement record. 


The emergence of momentum squeezing under the condition $\omega_\theta\ll\omega_m$ can be understood by examining the quantum estimation problem in the Fourier domain. 
The conditional spectral density exhibits a peak at $\omega_\theta$, which represents the effective resonance frequency of the estimated state (see Fig.~\ref{SqqSpp}). 
When $\omega_\theta\ll\omega_m$, the Wiener-filtered estimate of the mirror behaves as an \emph{effectively free particle}. 
In quantum mechanics, it is well known that the momentum of a free particle can be measured without disturbing the system, a property known as quantum non-demolition (QND). 
By analogy, we infer that, in the present system, the momentum can be estimated with higher precision than the position. 

As the coupling strength $g$ increases, the parameter $\xi$ approaches $\kappa/\Delta$. 
Moreover, for large $g$, the optical spring effect becomes significant, leading to an increase in the mechanical resonance frequency $\omega_m$. 
In this regime, the parameter $\eta$ can become sufficiently small for momentum squeezing to occur, as demonstrated in Fig.~\ref{thermal-noise}. 
Therefore, the optical spring effect, by enhancing the resonance frequency $\omega_m$, facilitates the condition $\omega_\theta\ll\omega_m$, thereby resulting in an effectively free-particle state. 

We now examine the condition for momentum squeezing in a quantitative manner. 
In the regime where the thermal noise is negligible compared to the radiation-pressure noise, $\eta\ll\xi$, and for $2\ll\xi$, one finds that the effective resonance frequency satisfies the approximate relation $(\omega_\theta/\omega_m)^4\simeq(\eta\xi+1)/\xi^2$. 
Using this approximation, the variances can be written approximately as 
\begin{align}
    \label{etaxiap}
    V_{qq} \simeq \sqrt{2\xi\sqrt{\eta\xi+1}},\; V_{pp} \simeq \sqrt{2\left(\eta+\frac{1}{\xi}\right)\sqrt{\eta\xi+1}}. 
\end{align}
In particular, when $\eta\xi<1$, these expressions reduce to 
\begin{align}
    \label{approx:etaxi<1}
    V_{qq} \simeq \sqrt{2\xi},\; V_{pp} \simeq \sqrt{\frac{2}{\xi}}, 
\end{align}
which clearly demonstrates the emergence of momentum squeezing for large $\xi$. 
On the other hand, when $\eta\xi>1$, we obtain 
\begin{align}
    \label{approx:etaxi>1}
    V_{qq} \simeq \sqrt{2\xi\sqrt{\eta\xi}},\; V_{pp} \simeq \sqrt{2\eta\sqrt{\eta\xi}}. 
\end{align}
The condition $\eta\xi > 1$ imposes an upper bound on the input power $P_\mathrm{in}$, which can be written as 
\begin{align}
    P_\mathrm{in} \lesssim \frac{m\ell^2\Omega^2\kappa^3}{8\omega_L\Delta} \frac{k_BT\Gamma\kappa}{\hbar\Omega^2\Delta}. 
\end{align}
In this regime, assuming $\Omega\ll\omega_m$, corresponding to a parameter region in which the optical spring effect is strong ($\Delta\neq0$), the momentum variance can be further approximated as 
\begin{align}
    \label{vppapprox}
    V_{pp} \sim \sqrt{\frac{\kappa^5}{4\Delta^2} \left(\frac{k_B T \Gamma m \ell^2}{\hbar P_\mathrm{in}\omega_c}\right)^{3/2}}. 
\end{align}
 From Eq.~\eqref{vppapprox}, the condition for momentum squeezing can then be approximately expressed as 
\begin{align}
    &\Bigl(\frac{T}{300\mathrm{K}}\Bigr)^{3/4}
    \Bigl(\frac{\Gamma/2\pi}{10^{-7}~\mathrm{Hz}}\Bigr)^{3/4}
    \Bigl(\frac{P_\mathrm{in}}{2\times10^{-3}~\mathrm{W}}\Bigr)^{-3/4}
    \Bigl(\frac{\kappa/2\pi}{10^{4}~\mathrm{Hz}}\Bigr)^{3/2}
    \Bigl(\frac{\Delta/\kappa}{0.05}\Bigr)^{-3/4}
    \nonumber\\
    &~~~~~~\quad\times
    \Bigl(\frac{\omega_c/2\pi}{2.8 \times 10^{14}~\mathrm{Hz}}\Bigr)^{-3/4}
    \Bigl(\frac{m}{100~\mathrm{mg}}\Bigr)^{3/4}
    \Bigl(\frac{\ell}{10~\mathrm{cm}}\Bigr)^{3/2}
    \simlt 1.
    \label{approximation}
\end{align}

We also note that momentum squeezing can occur even when $\Delta = 0$. 
In this case, no optical spring effect is present, and the condition for momentum squeezing differs from that given in Eq.~\eqref{approximation}. 

In the regime $\zeta \simeq \xi \sim \kappa/\Delta \gg 1$, we have $\alpha \simeq 2\Delta/\kappa$ and $\theta \simeq \alpha/2 \simeq \Delta/\kappa$ from Eq.~\eqref{deftheta}. 
The quantum state of the mirror cannot be inferred when the measurement rate vanishes, which occurs for $\lambda_\theta = 0$, yielding $\theta = \alpha$. 
The difference between this condition and the value of $\theta$ given by Eq.~\eqref{deftheta} is approximately $1/\xi \simeq \Delta/\kappa$ for $\Delta/\kappa \ll 1$, indicating that the two angles are close to each other in this limit, as illustrated by the solid curve and dashed curves in Fig.~\ref{VppContour}. 


\section{Gravity-induced entanglement}
\label{Gravity-induced entanglement}
We next consider the entanglement of a two-mode Gaussian state, which is characterized by the covariance matrix
\begin{align}
    \bm{V} \equiv
    \begin{pmatrix}
        \bm{V}_A & \bm{V}_{AB} \\
        \bm{V}_{AB}^\mathsf{T} & \bm{V}_B
    \end{pmatrix},
\end{align}
where $\bm{V}_A$ and $\bm{V}_B$ are the $2\times2$ covariance matrices of subsystems A and B, respectively, and $\bm{V}_{AB}$ represents the intersystem correlation matrix. 
The entanglement of a two-mode Gaussian state can be conveniently quantified by the logarithmic negativity, defined as $E_N = \max\{0,-\log_2\nu_-\}$, where $\nu_-^2 = (\Sigma - \sqrt{\Sigma^2 - 4\det\bm{V}})/2$ and $\Sigma = \det\bm{V}_A + \det\bm{V}_B - 2\det\bm{V}_{AB}$. 
According to the separability criterion for two-mode Gaussian states, the two subsystems are entangled if and only if $E_N > 0$ \cite{Giedke2001}. 

\begin{figure}[t]
    \centering
    \includegraphics[width=100mm]{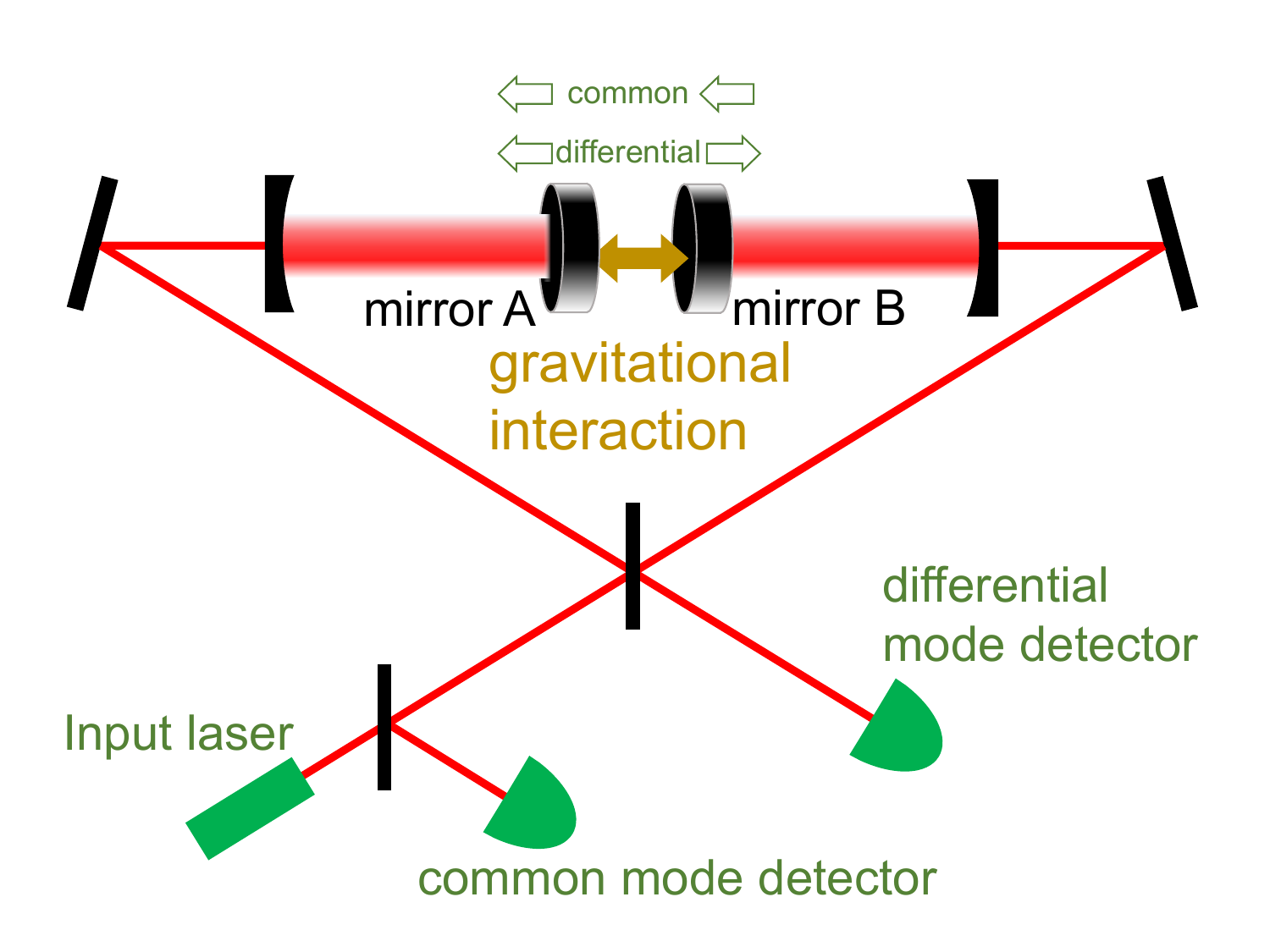}
    \caption{Configuration of the experimental setup. 
            We consider two cylindrical mirrors A and B with radius $R$ and thickness $h$. 
            The gravitational interaction affects only the differential mechanical mode associated with the relative displacement of the two mirrors. 
            }
    \label{setup}
\end{figure}

We now apply the results of the previous section to the detection of gravity-induced entanglement (GIE). 
In particular, we consider two cavity optomechanical systems, each consisting of a movable mirror, that interact via Newtonian gravity, as illustrated in Fig.~\ref{setup} (see also Refs.~\cite{Miao2020,Datta2021,Miki2024a,Miki2024b,Matsumoto2025,Hatakeyama2025}). 
The gravitational interaction is described by 
\begin{align}
    \hat{H}_\mathrm{int} = -\frac{\hbar\Omega\delta}{4}\hat{q}_-^2, 
\end{align}
where $\hat{q}_- = (\hat{q}_A - \hat{q}_B)/\sqrt{2}$ denotes the differential mechanical position mode. 
Here, $\hat{q}_{A}$ and $\hat{q}_{B}$ refer to the original mechanical position operators, prior to the rescaling introduced in Eq.~\eqref{rescaled quadratures}. 
The dimensionless parameter $\delta$ characterizes the gravitational coupling and is determined by the geometry and configuration of the mirrors, as well as by their material properties. 
Specifically, 
\begin{align}
    \delta = \frac{4G\rho\Lambda}{\Omega^2} 
\end{align}
depends on the gravitational constant $G$, the mechanical resonance frequency $\Omega$, the mass density $\rho$, and a geometric form factor $\Lambda$, which encodes the spatial mass distribution and the relative separation of the mirrors. 

In the present analysis, we assume a low mechanical frequency, $\Omega/2\pi = 10^{-3}\,\mathrm{Hz}$, and a high-density mirror material corresponding to a Pt--Au alloy with mass density $\rho = 20\,\mathrm{g/cm^{3}}$. 
The geometric form factor $\Lambda$ is evaluated following the assumptions and procedures described in Refs.~\cite{Matsumoto2025,Miao2020}, where a geometric constraint ensuring a physically realistic separation between the mirrors is explicitly imposed. 
For cylindrical mirrors with radius $R = 0.5h$ and thickness $h = 0.8L$, where $L$ denotes the center-of-mass separation, the form factor approaches an optimal value of order unity, $\Lambda \simeq 1$. 
Under these conditions, we obtain $\delta \simeq 0.14$. 

The gravity-induced entanglement (GIE) in this optomechanical setup originates from the asymmetry between the differential and common mechanical modes, which are characterized by $\Omega_+ = \Omega$ and $\Omega_- = \Omega\sqrt{1 - \delta}$, respectively. 
A stronger asymmetry between the common and differential modes allows for a larger amount of achievable entanglement. 
Momentum squeezing not only accentuates the difference between these two modes, but also—when the purity of the quantum state is sufficiently high—enhances the position uncertainty via the uncertainty principle. 
As a result, the spatial extent of the mechanical superposition is enlarged, thereby strengthening the GIE. 

\begin{figure}[h]
    \centering
    \includegraphics[width=100mm]{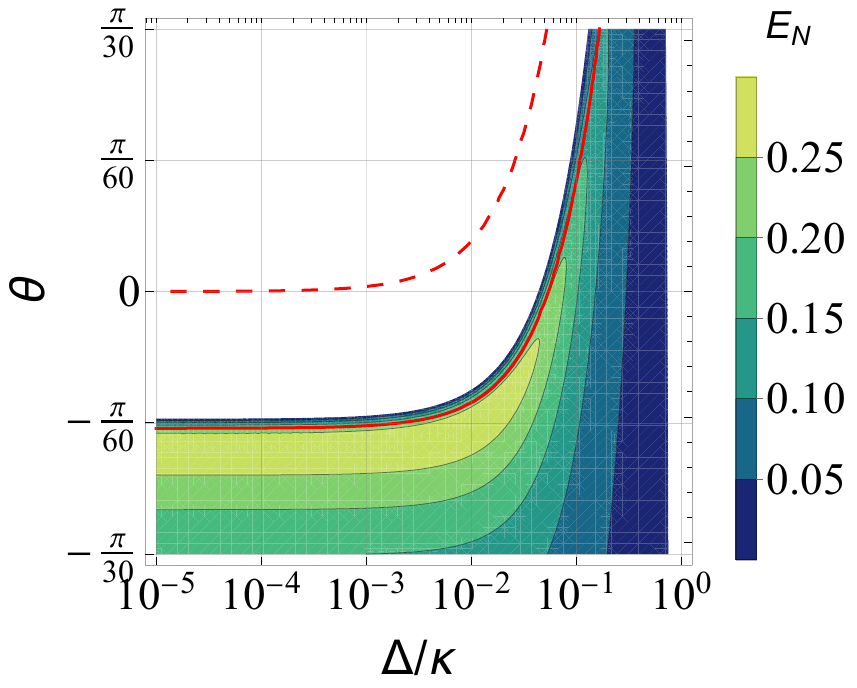}
    \caption{Same as Fig.~\ref{VppContour}, but showing a contour plot of the logarithmic negativity $E_N$. 
            We consider two optomechanical systems, as shown in Fig~\ref{setup}, each with the same parameters as those used in Fig.~\ref{SqqSpp}, except for $\theta$ and $\Delta$, and assume a gravitational coupling of $\delta\simeq0.14$. 
            }
    \label{Fig_Large Entanglement around p-squeeze}
\end{figure}

Fig.~\ref{Fig_Large Entanglement around p-squeeze} shows the contour plot of the logarithmic negativity $E_N$ as a function of the detuning $\Delta/\kappa$ and the homodyne angle $\theta$. 
Strong entanglement is predominantly observed in the parameter region where momentum squeezing occurs, as ensured by the condition $\omega_\theta \ll \omega_m$. 
This clearly indicates that the enhancement of entanglement originates from momentum squeezing. 

The condition for momentum squeezing given by Eq.~\eqref{deftheta} yields $\theta = \alpha - 1/\zeta_\pm \simeq \Delta/\kappa$ for $\zeta_\pm \gg 1$, where $\zeta_\pm$ is defined analogously to $\zeta$, with $\Omega$ replaced by $\Omega_\pm$. 
For the parameters considered in this work, the difference between $1/\zeta_+$ and $1/\zeta_-$ is negligible. 

The condition for generating gravity-induced entanglement (GIE) can be expressed in the form 
\begin{align}
    \label{encrr}
    2\Gamma(2n_\mathrm{th}+1) < \frac{4G\rho\Lambda}{\omega_m},
\end{align}
which can be interpreted as the requirement that the characteristic generation time of GIE be shorter than the mechanical dissipation timescale. 
The inequality~\eqref{encrr} can be rewritten as 
\begin{align}
    \label{ConditionE}
    {\left(\frac{T}{1\mathrm{K}}\right)}
    {\left(\frac{\Gamma/2\pi}{10^{-18}\mathrm{Hz}}\right)}
    {\left(\frac{20\mathrm{g/cm^{3}}}{\rho}\right)}
    {\left(\frac{1}{\Lambda}\right)}
    < 1.6. 
\end{align}

Following Ref.~\cite{CAVALLERI20103365}, the dissipation rate for a cylindrical-shaped mirror due to collisions with a background gas of particles with mass $m_\mathrm{atom}$, pressure $P$, and temperature $T$ can be written as 
\begin{align}
    \Gamma &= \frac{P\sqrt{m_\mathrm{atom}}}{m\sqrt{k_BT}}
    \pi R^2 \sqrt{\frac{8}{\pi}}
    \left(1+\frac{h}{2R}+\frac{\pi}{4}\right) \notag\\
    &= 2\pi \times 2.1\times 10^{-18} \times
    \left(\frac{P}{1\times 10^{-13}\mathrm{Pa}}\right)
    \left(\frac{100\mathrm{g}}{m}\right)^{1/3}
    \left(\frac{20\mathrm{g/cm^{3}}}{\rho}\right)^{2/3}
    \left(\frac{m_\mathrm{atom}}{1.7 \times 10^{-27}\mathrm{kg}}\right)^{1/2}
    \left(\frac{1\mathrm{K}}{T}\right)^{1/2}
    \mathrm{Hz}, 
\end{align}
where we have assumed a cylindrical geometry with radius $R = 0.5h$. 
Experimentally, pressures as low as $5\times10^{-15}\,\mathrm{Pa}$ at $4.2\,\mathrm{K}$ have been achieved using cryopumping \cite{Gabrielse2002}. 
Furthermore, V.~Baglin and R.~Kersevan reported that a pressure of $10^{-17}\,\mathrm{Pa}$ was achieved at $T=4.5\,\mathrm{K}$ in a high-energy physics experiment \cite{BaglinKersevan}. 
These results were obtained in ground-based experiments. 

The time required to generate gravity-induced entanglement scales as $\pi\Omega/(4G\rho)$, implying that operation at low mechanical frequencies, $\Omega/2\pi \sim 10^{-3}\,\mathrm{Hz}$, provides a significant advantage. 
The space environment offers a promising setting in which noise can be substantially reduced at such low frequencies; however, further detailed investigations are required to fully assess its feasibility (see, e.g., Ref.~\cite{Matsumoto2025}). 



\section{Conclusions}
\label{Conclusions}
We have investigated the conditional state of a mechanical mirror in an optomechanical system under continuous measurement, feedback control, and quantum filtering. 
In doing so, we have identified a parameter regime in which a highly momentum-squeezed state emerges. 
Our analysis shows that, by appropriately selecting the homodyne detection angle $\theta$ close to the condition given in Eq.~\eqref{deftheta}, such that $\omega_\theta \ll \omega_m$, it is possible to realize a momentum-squeezed regime characterized by $V_{pp}<1$, as indicated by the colored region in Fig.~\ref{VppContour}. 
This regime requires that the thermal noise be sufficiently small, as illustrated in Fig.~\ref{thermal-noise}. 
The phenomenon of the momentum squeezing via optimal filtering itself is expected to occur in a parameter regime that should be achievable in the relatively near future, as shown in Eq.~\eqref{approximation}. 

When momentum squeezing with $V_{pp}<1$ is achieved, the optimal filtering process effectively realizes a free-particle-like conditional state. 
This might be interpreted as enabling a backaction-evading measurement through momentum detection. 
Furthermore, we have demonstrated that this regime enhances the signal of gravity-induced entanglement proposed in Ref.~\cite{Miki2024b}. 
This enhancement arises because momentum squeezing increases the position uncertainty, thereby enlarging the spatial extent of the mechanical superposition via the uncertainty principle when the state purity is high. 
These findings are expected to play a significant role in advancing the quantum control of macroscopic systems and in enabling the generation of large gravity-induced entanglement for probing the quantum nature of gravity. 

We have also shown that our results do not depend on the feedback control parameters, even when feedback is included. 
Specifically, the noise introduced by the feedback can be canceled in the conditional state by applying optimal filtering. 
This has been demonstrated explicitly using the Kalman filtering approach in Appendix~\ref{Appendix:feedback}.

Finally, we emphasize that the primary goal of this work is not to claim a direct observation of a gravitationally induced entangled state. 
Rather, our aim is to show that conditional momentum squeezing—accessible via continuous measurement and optimal filtering—enhances the entanglement generation rate. 
This enhancement can be independently verified through measurable correlations, providing a realistic pathway toward experimental tests of the quantum nature of gravity. 

\begin{acknowledgements} 
We thank A.~Matsumura, S.~Iso, K.~Izumi, K.~Sakai, and N.~Matsumoto for helpful discussions. 
We also thank M.~Aspelmeyer and Y.~Chen for insightful discussions. 
K.Y. is supported by JSPS KAKENHI Grant No.~JP23H01175, and D.M. is supported by the JSPS Overseas Research Fellowships. 
\end{acknowledgements}

\appendix
\section{FEEDBACK CONTROL}\label{Appendix:feedback}
In this Appendix, we consider measurement-based feedback control and its effect on the conditional state in the presence of optimal filtering. 
The key result is obtained by designing the estimator sequentially so that it properly accounts for the non-Markovian noise generated by the feedback process. 

In Eq.~\eqref{Ffbt}, the term $F_\mathrm{fb}(t)$ is introduced to represent a feedback force applied to the mechanical system. 
This feedback force depends on the measurement record passed through a high-pass filter and can be written explicitly as 
\begin{align}
    F_\mathrm{fb}(t)&=-\int^{t}_{-\infty}G(t-s)\frac{\hat{I}_\theta(t)}{c_\theta}ds\nonumber\\
    &=-\frac{G_0\omega_\mathrm{fb}}{c_\theta}\hat{I}_\theta(t)
    +\frac{G_0\omega_\mathrm{fb}^2}{c_\theta}\int^{\infty}_0\Theta(s)e^{-\omega_\mathrm{fb}s}\hat{I}_\theta(t-s)ds,
    \label{Ffb}
\end{align}
where $G(t)=G_0\frac{d}{dt}\big(\Theta(t)\omega_\mathrm{fb}e^{-\omega_\mathrm{fb}t}\big)$ is a standard derivative high-pass filter \cite{Genes2008}. 
Here, $G_0>0$ denotes the feedback gain, $\omega_\mathrm{fb}^{-1}$ is the characteristic time delay of the feedback loop, and $\Theta(t)$ is the Heaviside step function. 
The second term on the right-hand side of Eq.~\eqref{Ffb} represents the noise introduced by the feedback control, which exhibits a memory effect. 

We now consider the estimation of the mechanical motion based on measurements of a general output field. 
Kalman filtering \cite{Kalmanf} and causal Wiener filtering \cite{Wienerf} provide essentially equivalent quantum estimation schemes in the steady-state regime. 
The distinction between these two approaches lies primarily in their domain of implementation: the time domain for Kalman filtering and the frequency domain for Wiener filtering. 

We first demonstrate that the contribution of the feedback does not appear in the conditional state dynamics when Kalman filtering is applied in the presence of feedback (FB). 
To this end, we rewrite the Langevin equations for the mechanical operators $\bm{r}(t)=(\hat{q}(t), \hat{p}(t))^{\top}$ in matrix form as 
\begin{align}
    \frac{d\bm{r}(t)}{dt}
    &=\bm{Ar}(t)+\bm{b}\hat{w}(t)
    -\frac{G_0\omega_\mathrm{fb}}{c_\theta}\bm{b}\big(\bm{Cr}(t)+\bm{b}\hat{v}_\theta(t)\big)
    +\bm{b}u_\mathrm{m}(t)\nonumber\\
    &=\Big(\bm{A}-\frac{G_0\omega_\mathrm{fb}}{c_\theta}\bm{bC}\Big)\bm{r}(t)
    +\bm{b}\hat{w}(t)
    -\frac{G_0\omega_\mathrm{fb}}{c_\theta}\bm{b}\hat{v}_\theta(t)
    +\bm{b}u_\mathrm{m}(t),
    \label{bare}
\end{align}
where $\hat{w}(t)$, $\hat{v}_\theta(t)$, and $c_\theta$ are defined in 
Eqs.~\eqref{defw}, \eqref{defv}, and \eqref{deftheta}, respectively. 
We have also introduced 
\begin{align}
    &\bm{A}=
    \begin{pmatrix}
     0 & \omega_m \\
    -\omega_m & -\Gamma
    \end{pmatrix}, \quad
    \bm{b}=
    \begin{pmatrix}
     0 \\
     1
    \end{pmatrix}, \quad
    \bm{C}=
    \begin{pmatrix}
     c_\theta & 0
    \end{pmatrix},\\
    &u_\mathrm{m}(t)=\frac{G_0\omega_\mathrm{fb}^2}{c_\theta}
    \int^{\infty}_0\Theta(s)e^{-\omega_\mathrm{fb}s}\hat{I}_\theta(t-s)ds.
\end{align}

The first and second moments of the mirror noise $\hat{w}(t)$ and the shot noise $\hat{v}_\theta(t)$ satisfy 
\begin{align}
    \langle \hat{w}(t)\rangle &= \langle \hat{v}_\theta(t)\rangle = 0, \nonumber\\
    \frac{1}{2}\langle\{ \hat{w}(t),\hat{w}(t')\}\rangle
    &=\Big(2\Gamma(2n_\mathrm{th}+1)
    +\frac{16g_m^2\kappa}{\kappa^2+4\Delta^2}\Big)\delta(t-t')
    :=\bar{n}\delta(t-t'),\\
    \frac{1}{2}\langle \{\hat{v}_\theta(t), \hat{v}_\theta(t')\}\rangle
    &=(2N_\mathrm{th}+1)\delta(t-t')
    :=M_\theta\delta(t-t'),\\
    \frac{1}{2}\langle \{\hat{w}(t),\hat{v}_\theta(t')\}\rangle
    &=\Big(
    \frac{4g_m\kappa}{\sqrt{\kappa^2+4\Delta^2}}
    \cos\!\Big(\theta-\arctan\frac{2\Delta}{\kappa}\Big)
    \Big)\delta(t-t')
    :=L_\theta\delta(t-t').
\end{align}
Here, $\bar{n}=\langle \hat{w}^2\rangle$ represents the combined thermal and radiation-pressure noise acting on the mirror, while $M_\theta=\langle \hat{v}_\theta^2\rangle$ characterizes the shot noise associated with the homodyne angle $\theta$. 
For phase-quadrature measurement with $\Delta=0$, the radiation-pressure noise originates from the amplitude quadrature, whereas the shot noise arises from the phase quadrature. 
In this case, the two noise sources are uncorrelated, yielding $L_\theta=0$ when $\theta=0$. 

We now focus on quantum estimation in the time domain and briefly review why the conditional  variances are independent of the feedback control. 
Here, we adopt the Kalman filter as a framework for quantum optimal estimation. 
Although the Kalman formalism assumes white noise, it provides the standard time-domain approach to optimal estimation. 
Importantly, the steady-state conditional variances obtained using the Kalman filter coincide with those derived from the causal Wiener filter employed in the main text. 

The Kalman filter is constructed by designing an optimal estimator based on the measurement record $\hat{I}_\theta(t)$. 
Even when the system exhibits a measurement-induced memory effect (i.e., non-Markovianity), the estimator can be designed to cancel this effect. 
This reflects the fact that, under measurement-based feedback, the experimenter has complete knowledge of when and how the feedback force is applied. 
Accordingly, the same non-Markovian noise is injected into the estimator, ensuring that it cancels exactly in the discrepancy between the system dynamics and the estimator $\overrightarrow{\bm{r}}(t)=(\overrightarrow{q}(t), \overrightarrow{p}(t))^{\top}$: 
\begin{align}
    \frac{d\overrightarrow{\bm{r}}(t)}{dt}
    =\Big(\bm{A}-\frac{G_0\omega_\mathrm{fb}}{c_\theta}\bm{bC}\Big)\overrightarrow{\bm{r}}(t)
    +\bm{b}u_m(t)
    +\bm{K}\big(\hat{I}_\theta(t)-\bm{C}\overrightarrow{\bm{r}}(t)\big).
    \label{estimator}
\end{align}
Here, $\bm{K}$ denotes the Kalman gain \cite{Zhang2018,Matsumoto}, which is chosen to minimize the estimation error $d\bm{e}(t):=d\bm{r}(t)-d\overrightarrow{\bm{r}}(t)$. 
The feedback-induced noise term $\bm{b}u_m(t)$ is included so as to cancel the corresponding term in the bare dynamics~\eqref{bare}. 

Under this choice of estimator, the error dynamics can be written as 
\begin{align}
    \frac{d\bm{e}(t)}{dt}
    =\Big(\Big(\bm{A}-\frac{G_0\omega_\mathrm{fb}}{c_\theta}\bm{bC}\Big)-\bm{KC}\Big)\bm{e}(t)
    +\bm{b}\hat{w}(t)
    -\frac{G_0\omega_\mathrm{fb}}{c_\theta}\bm{b}\hat{v}_\theta(t)
    -\bm{K}\hat{v}_\theta(t),
\end{align}
which contains only Gaussian noise terms. 

The time evolution of the second moment of the causal conditional state, $V(t)=\langle \bm{e}(t)\bm{e}(t)^{\top}\rangle$, can then be written as 
\begin{align}
    \dot{\bm{V}}(t)
    &=\Big(\Big(\bm{A}-\frac{G_0\omega_\mathrm{fb}}{c_\theta}\bm{bC}\Big)-\bm{KC}\Big)\bm{V}(t)
    +\bm{V}(t)\Big(\Big(\bm{A}-\frac{G_0\omega_\mathrm{fb}}{c_\theta}\bm{bC}\Big)-\bm{KC}\Big)^{\top}\nonumber\\
    &+\bm{b}\bar{n}\bm{b}^{\top}
    +\Big(\frac{G_0\omega_\mathrm{fb}}{c_\theta}\Big)^2\bm{b}M_\theta\bm{b}^{\top}
    +\bm{K}M_\theta\bm{K}^{\top}
    -\frac{2G_0\omega_\mathrm{fb}}{c_\theta}\bm{b}L_\theta\bm{b}^{\top}
    -\bm{b}L_\theta\bm{K}^{\top}
    -\bm{K}L_\theta\bm{b}^{\top}
    \label{RiccatiEq}
\end{align}
which can be rewritten in a compact form as 
\begin{align}
    \dot{\bm{V}}(t)
    =(\bm{A}'-\bm{KC})\bm{V}(t)
    +\bm{V}(t)(\bm{A}'-\bm{KC})^{\top}
    +\bm{b}\bar{n}_\mathrm{eff}\bm{b}^{\top}
    +\bm{K}M_\theta\bm{K}^{\top}
    -\bm{b}L_\mathrm{eff}\bm{K}^{\top}
    -\bm{K}L_\mathrm{eff}\bm{b}^{\top},
\end{align}
with 
\begin{align}
    \bm{A}'&:=\bm{A}-\frac{G_0\omega_\mathrm{fb}}{c_\theta}\bm{bC},\nonumber\\
    \bar{n}_\mathrm{eff}
    &:=\bar{n}-\frac{2G_0\omega_\mathrm{fb}}{c_\theta}L_\theta
    +\Big(\frac{G_0\omega_\mathrm{fb}}{c_\theta}\Big)^2M_\theta,\quad
    L_\mathrm{eff}:=L_\theta-\frac{G_0\omega_\mathrm{fb}}{c_\theta}M_\theta.
\end{align}

Since Eq.~\eqref{RiccatiEq} is formally equivalent to the forward equations derived in Refs.~\cite{Matsumoto,Zhang2018}, the optimal Kalman gain is given by $\bm{K} = (\bm{V}(t)\bm{C}^{\top} + \bm{b}L_\mathrm{eff})M_\theta^{-1}$. 
Substituting this expression, the evolution equation becomes 
\begin{align}
    \dot{\bm{V}}(t)
    =\bm{A'V}(t)+\bm{V}(t)\bm{A'}^{\top}
    +\bm{b}\bar{n}_\mathrm{eff}\bm{b}^{\top}
    -(\bm{V}(t)\bm{C}^{\top}+\bm{b}L_\mathrm{eff})
    M_\theta^{-1}
    (\bm{V}(t)\bm{C}^{\top}+\bm{b}L_\mathrm{eff})^{\top}.
    \label{Vdot_feedback}
\end{align}

By examining each term on the right-hand side in detail, one finds that all feedback-related contributions cancel exactly. 
As a result, the evolution equation reduces to 
\begin{align}
    \dot{\bm{V}}(t)
    =\bm{AV}(t)+\bm{V}(t)\bm{A}^{\top}
    +\bm{b}\bar{n}\bm{b}^{\top}
    -(\bm{V}(t)\bm{C}^{\top}\bm{b}+L_\theta)
    M_\theta^{-1}
    (\bm{V}(t)\bm{C}^{\top}\bm{b}+L_\theta)^{\top}.
\end{align}
This expression is identical to the result obtained by performing Kalman filtering without feedback (i.e., $G_0=0$). 
The steady-state solution for the conditional variances, obtained by setting the left-hand side to zero, is therefore the same as that reported in Refs.~\cite{Matsumoto,Miki2024b}:
\begin{align}
    &V_{qq}=\frac{\gamma_\theta-\Gamma}{\lambda_\theta},\quad
    V_{pp}=V_{qq}\Big(\Big(\frac{\omega_\theta}{\omega_m}\Big)^2
    -\frac{\Gamma(\gamma_\theta-\Gamma)}{2\omega_m^2}\Big),\quad
    V_{qp}=V_{qq}\frac{\gamma_\theta-\Gamma}{2\omega_m^2}.
    \label{CV}
\end{align}
Since these conditional variances contain no contribution from the feedback parameters $(G_0,\omega_\mathrm{fb})$, we conclude that measurement-based feedback control using a high-pass filter does not affect the conditional variances. 

\begingroup
\raggedright
\bibliography{main_letter_v5}
\endgroup

\end{document}